\documentclass{amsproc}

\usepackage{amssymb}

\usepackage{latexsym, MnSymbol, amscd}

\usepackage{color, xcolor}

\newcommand{\Si}{ {\Sigma} }

\newcommand{\bC}{\mathbb{C}}
\newcommand{\bD}{\mathbb{D}}

\newcommand{\bF}{\mathbb{F}}
\newcommand{\bH}{\mathbb{H}}

\newcommand{\bP}{\mathbb{P}}
\newcommand{\bQ}{\mathbb{Q}}
\newcommand{\bR}{\mathbb{R}}

\newcommand{\bZ}{\mathbb{Z}}

\newcommand{\cC}{\mathcal{C}}

\newcommand{\cF}{\mathcal{F}}

\newcommand{\cL}{\mathcal{L}}
\newcommand{\cM}{\mathcal{M}}
\newcommand{\cO}{\mathcal{O}}
\newcommand{\cP}{\mathcal{P}}

\newcommand{\cT}{\mathcal{T}}

\newcommand{\vir}{ {\mathrm{vir}} }
\newcommand{\Sym}{ \mathrm{Sym} }

\newcommand{\tcM}{\widetilde{\cM}}
\newcommand{\tcT}{\widetilde{\cT}}

\newcommand{\Mbar}{\overline{\cM}}
\newcommand{\dbar}{\bar{\partial}} 

\newtheorem{theorem}{Theorem}[section]

\theoremstyle{definition}

\newtheorem{example}[theorem]{Example}

\theoremstyle{remark}

\numberwithin{equation}{section}

\begin{document}

\title[Holomorphic Anomaly Equations]{A Lecture on 
Holomorphic Anomaly Equations and Extended Holomorphic Anomaly Equations}


\author{Chiu-Chu Melissa Liu}
\address{Department of Mathematics, Columbia University, 2990 Broadway, New York, NY 10027, USA}
\curraddr{}
\email{ccliu@math.columbia.edu}
\thanks{}


\date{May 10, 2019}


\maketitle

\section{Mirror Symmetry for Compact Calabi-Yau threefolds}

Mirror symmetry relates the A-model on a compact Calabi-Yau threefold $X$, defined in terms of the symplectic structure on $X$, 
to the B-model on a mirror compact Calabi-Yau threefold $Y$, defined in terms of the complex structure on $Y$,
$$
\textup{A-model }(X,\omega) \ \stackrel{\textup{Mirror Symmetry}}{\longleftrightarrow} \ \textup{B-model }(Y,\Omega),
$$
where $\omega$ is a Ricci flat K\"{a}hler form on $X$ and $\Omega$ is a holomorphic volume form (i.e. a nowhere vanishing holomorphic 3-form) on $Y$.

The Hodge diamond of $X$ is of the form
$$
\begin{array}{ccccccc}
&&&  1 &&& \\
&& 0 && 0 && \\
& 0 && h^{2,2}(X) && 0 \\
1 \quad &  & h^{2,1}(X) && h^{1,2}(X) &&  \quad 1 \\
& 0 && h^{1,1}(X) && 0\\
 && 0 && 0 && \\
 &&& 1. &&&
\end{array}
$$
The Hodge diamond of $Y$ is of the same form. Let
$$
\kappa:=h^{1,1}(X)=h^{2,2}(X) = \dim_{\bC}H^2(X;\bC) =\dim_{\bC}\cM_A, 
$$
where $\cM_A$ is the complexified K\"{a}hler moduli of $X$. Then
$$
\kappa= h^{2,1}(Y)=h^{1,2}(Y)=\dim_\bC H^1(Y,T_Y) =\dim_\bC \cM_B, 
$$
where $\cM_B$ is the moduli of complex structures on $Y$.

\section{A-model on compact Calabi-Yau threefolds}

\subsection{A-model topological closed strings: Gromov-Witten invariants}
Given a non-negative integer $g$ and an effective curve class $\beta\in H_2(X;\bZ)$, let $\Mbar_g(X,\beta)$ be the moduli space of
genus $g$ degree $\beta$ stable maps to $X$ (with no marked points). Note that $\Mbar_g(X,\beta)$ is empty if 
$(g,\beta) =  (0,0)$ or $(1,0)$. The moduli $\Mbar_g(X,\beta)$ is a proper Deligne-Mumford stack equipped with a
virtual fundamental class
$$
[\Mbar_g(X,\beta)]^\vir\in H_0(\Mbar_g(X,\beta);\bQ).
$$
The genus $g$, degree $\beta$ Gromov-Witten invariant of $X$ is defined by
$$
N_{g,\beta}^X:= \int_{[\Mbar_g(X,\beta)]^\vir} 1\in \bQ
$$
when $(g,\beta)\neq (0,0),(1,0)$. The integral sign in the above equation stands for the natural pairing between
$[\Mbar_g(X,\beta)]^\vir\in H_0(\Mbar_g(X,\beta);\bQ)$ and $1\in H^0(\Mbar_g(X,\beta);\bQ)$. For $g\geq 2$, we have
$$
\Mbar_g(X,0) =\Mbar_g\times X,
$$
and
$$
N_{g,0}^X = \frac{(-1)^g}{2} \int_{\Mbar_g}\lambda_{g-1}^3\int_X c_3(X) 
=  \frac{(-1)^g |B_{2g}|\cdot |B_{2g-2}|}{4g(2g-2)\cdot (2g-2)!} \int_X c_3(X), 
$$
where $B_{2m}$ are Bernoulli numbers. 

We choose a basis $H_1,\ldots,H_k$ of $H^{1,1}(X)=H^2(X;\bC)$ such that each $H_i$ is in $H^2(X;\bZ)$ and also in the nef cone
(which is the closure of the K\"{a}hler cone of $X$). A complexified K\"{a}hler class is of the form
$$
t = \sum_{i=1}^\kappa t^i H_i,
$$
where $t^1,\ldots,t^\kappa\in \bC$ are complexified K\"{a}hler parameters. 
The genus $g$ Gromov-Witten potential of $X$, $F_g^X$, is a generating function of genus 
$g$ Gromov-Witten invariants of $X$:
$$
F_g^X(t) := \begin{cases}
\displaystyle{\frac{1}{6}\int_X t^3 + \sum_{\beta\neq 0} N_{0,\beta}^X \exp(\int_\beta t)}, & g=0;\\
\displaystyle{-\frac{1}{24}\int_X c_2(X)t+ \sum_{\beta\neq 0} N_{1,\beta}^X \exp(\int_\beta t) }, & g=1;\\
\displaystyle{ N_{g,0}^X + \sum_{\beta\neq 0} N_{g,\beta}^X \exp(\int_\beta t)}, & g\geq 2.
\end{cases}
$$
The sum $\displaystyle{\sum_{\beta\neq 0}}$ is over all non-zero effective classes. We may write
$$
F_g^X = F_g^{X,\textup{classical}} +F_g^{X,\textup{quantum}},
$$
where
$$
F_g^{X,\textup{quantum}} =\sum_{\beta\neq 0} N_{g,\beta}^X \exp(\int_\beta t)
$$
is the contribution from non-constant genus $g$ stable maps to $X$. We have
$$
\exp(\int_\beta t) = Q_1^{d_1}\cdots Q_\kappa^{d_\kappa},
$$
where $Q_i = \exp(t^i)$ and $d_i =  \int_\beta H_i\in \bZ_{\geq 0}$. So
$F_g^{X,\mathrm{quantum}} $ is a formal power series in $Q_1,\ldots, Q_\kappa$ with rational coefficients; it tends to  zero at the large radius limit
$Q_i\to 0$:
$$
\lim_{Q_i\to 0} F_g^{X,\mathrm{quantum}}(t) =0.
$$
We have
\begin{eqnarray*}
&& \frac{\partial ^3 F_0^X}{\partial t^i\partial t^j\partial t^k} = (H_i\star H_j, H_k)\\
&=& \int_X H_i\cup H_j\cup H_k  + \sum_{\beta\neq 0} \Big(\int_\beta H_i \int_\beta H_j  \int_\beta H_k \Big)N_{0,\beta}^X \exp(\int_\beta t),
\end{eqnarray*}
where $\star$ is the quantum product, where $\cup$ is the classical cup product, and 
where $(\ ,\ )$ is the Poincar\'{e} pairing on $H^*(X)$.

\subsection{A-model topological open strings: open Gromov-Witten invariants}
Let $L\subset X$ be a closed oriented Lagrangian submanifold. Then the tangent bundle of $L$ is trivial -- recall that the tangent bundle of any orientable 3-manifold is trivial. In this paper, we further assume that $L$ is a rational homology 3-sphere:
$$
H_*(L;\bQ)= H_*(S^3;\bQ). 
$$
Then the map $H_2(X;\bQ)\to H_2(X,L;\bQ)$ is an isomorphism, so $H_2(X;\bZ)\to H_2(X,L;\bZ)$ has finite kernel and cokernel.  
Since $H_1(L;\bZ)$ is torsion and $H_0(L;\bZ)=\bZ$, we have
$H^1(L;\bZ)=0$ by the universal coefficient theorem. In particular, the Maslov class $\mu(L)\in H^1(L;\bZ)$ (which is defined for any Lagrangian submanifold in a Calabi-Yau manifold) is zero.   Given a pair $(g,h)$, where $g$ is a nonnegative integer 
and $h$ is a positive integer, let $\Mbar_{(g,h)}(X,L,\beta)$ be the stable compactification of the moduli $\cM_{g,h}(X,L,\beta)$, 
parametrizing holomorphic maps $u: (\Si,\partial \Si)\to (X,L)$, where $\Si$ is a bordered Riemann surface with $g$ handles and $h$ holes, 
and $u_*[\Sigma]= \beta  \in H_2(X,L;\bZ)$, where the domain $\Si$ is oriented by its complex structure.   Then $\Mbar_{g,h}(X,L,\beta)$ is a (usually singular) orbifold whose virtual dimension is zero.  In some cases it is possible to define 
a virtual number $N_{(g,h),\beta}^{X,L}$ of points in $\Mbar_{g,h}(X,L,\beta)$; in general, $N_{(g,h),\beta}^{X,L}$ is a rational number (instead of an integer) due to the existence of orbifold points.  
We define generating functions of open Gromov-Witten invariants of the pair $(X,L)$ by 
$$
F_{(g,h)}^{X,L}(t) = \sum_{\beta\neq 0} N^{X,L}_{(g,h),\beta} \exp(\int_\beta t),
$$
where the sum is over nonzero relative homology classes $\beta\in H_2(X,L;\bZ)$. By assumption, for any $\beta\in H_2(X,L;\bZ)$ there
exists a positive integer $r$ such that $r\beta$ lies in the image of $H_2(X;\bZ)\to H_2(X,L;\bZ)$, so $F_{(g,h)}^{X,L}(t)$
is a formal power series in $Q_1^{1/r_1},\ldots, Q_\kappa ^{1/r_\kappa}$ for some positive   integers  $r_1,\ldots, r_\kappa$, and it tends to zero at the large radius limit:
$$
\lim_{Q_i\to 0} F_{(g,h)}^{X,L}(t) =0.
$$

\begin{example}
Let $X$ be  a quintic Calabi-Yau threefold with real coefficients,  and let $L$ be the real quintic. Then $L$ is a Lagrangian submanifold of $X$, which is
diffeomorphic to $\bR\bP^3$, so it is orientable and is a rational homology sphere. 
The group homomorphism 
$$
H_2(X;\bZ)= \bZ \longrightarrow H_2(X,L;\bZ)=\bZ
$$
is injective with cokernel $\bZ/2\bZ$. In this case $\kappa=1$, and
$$
F_g^{X,\textup{quantum}} \in \bQ[\![ Q ]\!],\quad F_{(g,h)}^{X,L}\in \bQ[\![Q^{1/2}]\!],
$$ 
where $Q=Q_1$. 
\end{example}

\section{Preliminaries on moduli of complex structures}

\subsection{The complex moduli and the vacuum line bundle} \label{sec:complex}
Recall that $\cM_B$ is the moduli space of complex structures on $Y$, and that $\dim_\bC\cM_B =\kappa$. 
Let $q=(q_1,\ldots,q_\kappa)$ be the local holomorphic coordinates
on $\cM_B$ such that $q = 0$ corresponds to a maximal unipotent monodromy point in the boundary of a (partial) compactification 
$\Mbar_B$ of $\cM_B$. (In particular, $q=0$ is in $\Mbar_B\setminus \cM_B$.) 
Let  $\cL^\vee \to \cM_B$ be the complex line bundle over $\cM_B$ whose fiber over $q\in \cM_B$ is
$\cL^\vee_q= H^0(Y_q,\Omega^3_{Y_q})\cong \bC$; then $\cL^\vee$ is a holomorphic line bundle over $\cM_B$, and its dual $\cL$ 
is the vacuum line bundle in the physics literature such as \cite{S} and \cite{BCOV}. 
The extended moduli space $\tcM_B$ is the total space of the frame bundle of $\cL^\vee$; it parametrizes
pairs $(Y_q,\Omega)$, where $Y_q$ corresponds to a point  $q\in \cM_B$ and $\Omega$ is a nonzero holomorphic 3-form on $Y_q$. So
$p: \tcM_B \to \cM_B$ is a principal $\bC^*$-bundle, and $\dim_\bC \tcM_B = \kappa+1$.
  
\subsection{The Torelli space}
Let $H$ denote the rank 2 lattice $\bZ\oplus \bZ$ equipped with the symplectic form $\Big(\begin{array}{cc} 0&1\\ -1& 0\end{array}\Big)$.
The Torelli space of $Y$ is the moduli of the marked Calabi-Yau threefold $(Y_q,\gamma)$, where $Y_q$ corresponds to a point $q\in \cM_B$ and
the marking $\gamma$ is an isometry from $H^{\oplus (\kappa+1)}$ to $H^3(X_q;\bZ)/\mathrm{Tor}$. Forgetting the marking
$\gamma$ defines a covering map $\pi: \cT\to \cM_B$, which is a principal $Sp(2\kappa+2;\bZ)$-bundle. Let $\tcT$ be the fiber product:
$$
\begin{CD}
\tcT @>>>  \tcM_B \\
@VVV  @VV{p}V\\
\cT @>{\pi}>> \cM_B.
\end{CD}
$$
Then $\tcT \to \cT$ is a principal $\bC^*$-bundle that is the frame bundle of $\pi^*\cL^\vee$, 
and $\tcT\to\tcM_B$ is a covering map that is a principal
$Sp(2\kappa+2;\bZ)$ bundle.

\subsection{The Hodge bundle $\bH_\bC$ and the Gauss-Manin connection}
Let $\bH_\bZ$ be the local system of lattices on $\cM_B$ whose fiber over $q\in \cM_B$ is $H^3(Y_q;\bZ)\cong \bZ^{2k+2}$. 
Then $\bH_\bR = \bH_\bZ\otimes_{\bZ}\bR$ (resp. $\bH_\bC= \bH_\bZ\otimes_\bZ \bC$) is a flat real (resp. complex) vector bundle 
of rank $2\kappa+2$ whose  fiber at $q\in \cM_B$ is $H^3(Y_q;\bR)$ (resp. $H^3(Y_q;\bC)$); the flat connection is known as the Gauss-Manin connection.

More explicitly, let
$$
\nabla: \Omega^0(\cM_B, \bH_\bC) = C^\infty(\cM_B, \bH_\bC) \longrightarrow \Omega^1(\cM_B, \bH_\bC) = 
C^\infty(\cM_B, T^*M_B\otimes \bH_\bC)
$$
be the Gauss-Manin connection. Let $U$ be an open subset on $\cM_B$ such that $\bH_\bZ|_U$ is trivial; we choose a trivialization of
$\bH_\bZ|_U$, or equivalently, a symplectic basis $\{ \alpha_i,\beta^i: i=0,1,\ldots, \kappa\}$ of $H^3(Y_q;\bZ) \cong H^{\oplus (\kappa+1)}$ 
for $q\in U$:
$$
\int_{Y_q} \alpha_i \cup \alpha_j = \int_{Y_q} \beta^i\cup \beta^j=0,\quad
\int_{Y_q} \alpha_i\cup \beta^j = -\int_{Y_q} \beta^j \cup \alpha_i =\delta_{ij}.
$$
Then $\{\alpha_i, \beta^i\}$ is a frame of $\bH_\bC|_U$ and 
$$
\nabla \alpha_i = \nabla \beta^i =0,\quad i=0,1,\ldots, \kappa.
$$ 
Any $C^\infty$ section $s \in C^\infty(U,\bH_\bC)$ is of the form
\begin{equation}\label{eqn:s}
s =\sum_{i=0}^\kappa (a^i \alpha_i + b_i \beta^i), 
\end{equation}
where $a^i, b_i$ are complex-valued $C^\infty$ functions on $U$. Then
$$
\nabla s  = \sum_{i=0}^\kappa (da^i \alpha_i + db_i \beta^i),
$$
where $da^i, db_i \in \Omega^1(U,\bC)$ are $C^\infty$ 1-forms on $U$.

We may write
$$
\nabla = \nabla' + \nabla'',
$$
where $\nabla':\Omega^0(\cM_B,\bH_\bC) \to \Omega^{1,0}(\cM_B,\bH_\bC)$ is a $(1,0)$-connection on the $C^\infty$ complex vector bundle
$\bH_\bC$  and where $\nabla'':\Omega^0(\cM_B,\bH_\bC)\to \Omega^{0,1}(\cM_B,\bH_\bC)$ is a $(0,1)$-connection on $\bH_\bC$. The $(0,1)$-connection
$\nabla''$ defines a holomorphic structure on $\bH_\bC$: a section $s \in C^\infty(U,\bH_\bC)$ is holomorphic iff 
$\nabla'' s =0$ iff
$$
s =  \sum_{i=0}^\kappa (a^i \alpha_i + b_i \beta^i),
$$
where $a^i, b_i$ are holomorphic functions on $U$. 

\subsection{Hodge filtration and the holomorphic polarization}
We have  the  Hodge filtration
$$
0\subset \bF^3\subset \bF^2 \subset \bF^1\subset \bF^0 =\bH_\bC,
$$
where 
$$
\bF^i_q = \bigoplus_{p\geq i} H^{p,3-p}(Y_q;\bC). 
$$
The complex vector bundles $\bF^1, \bF^2, \bF^3$ are (non-flat) holomorphic subbundles of $\bF^0=\bH_\bC$ 
of ranks $2\kappa+1$, $\kappa+1$, $1$, respectively. In particular, $\bF^3=\cL^\vee$ is the dual of the vacuum line bundle. 
We also have
$$
\bH_\bC = \bF^2 \oplus \overline{\bF^2},
$$
where
$$
\bF^2_q = H^{3,0}(Y_q;\bC)\oplus H^{2,1}(Y_q;\bC),\quad \overline{\bF^2_q} = H^{1,2}(Y_q;\bC) \oplus H^{0,3}(Y_q;\bC).
$$
For each $q\in U$, 
$$
H^3(Y_q;\bC)  = \bF^2_q\oplus \overline{\bF^2_q}
$$
is the holomorphic polarization of the complex symplectic space $H^3(Y_q;\bC)$.

\subsection{Real polarization}
We choose a symplectic basis $\{ A^i, B_i: i=0,1,\ldots,\kappa\}$ of $(H_3(Y;\bZ),\cap)$, where $\cap$ is the intersection form, and let 
$\{\alpha_i, \beta^i: i=0,1,\ldots, \kappa\}$ be the dual symplectic basis of $(H^3(Y;\bZ), Q)$, where
$$
Q(\alpha, \beta) = \int_Y \alpha\cup \beta \in \bZ 
\quad \textup{ for } \alpha,\beta \in H^3(Y;\bZ). 
$$
We have
$$
\delta^i_j = A^i\cap B_j = - B_j \cap A^i = \int_Y \alpha_i \cup \beta^j= -\int_Y \beta ^j\cup \alpha_i = \int_{A^i} \alpha_j =
\int_{B_j}\beta^i.
$$
$$
0 =A^i\cap A^j = B_i \cup B_j = \int_Y \alpha_i\cup \alpha_j =\int_Y \beta^i \cup \beta^j =\int_{A^i}\beta^j = \int_{B_j} \alpha_i.
$$
For any $\phi \in H^3(X;\bR)$ we have
$$
\phi  =\sum_{i=0}^\kappa x^i  \alpha_i +\sum_{i=0}^\kappa p_i \beta^i,
$$
where 
$$
x^i=\int_{A^i}\phi,\quad p_i =\int_{B_i}\phi.
$$
Then $\{ x^i, p_i: i=0,1,\ldots, k\}$ are Darboux coordinates of the real linear symplectic space $(H^3(Y;\bR),Q)$. The linear symplectic form
$$
\sum_{i=0}^k dx^i\wedge dp_i
$$
on $H^3(Y;\bR)$ is independent of the choice of a symplectic basis. 
The integral symplectic basis $\{\alpha_i ,\beta^i : i=0,1,\ldots,\kappa\}$ extends to a flat frame
of $\bH_\bR$ on an open neighborhood $U$ of $q_0$ in $\cM_B$, and $\{ x^i, p_i : i=0,1,\ldots, \kappa\}$ are flat fiber coordinates
of $\bH_\bR|_U \cong U\times \bR^{2\kappa +2}$.

The Gauss-Manin connection is compatible with the symplectic structure $Q$ on $\bH_\bC$: given two $C^\infty$ sections
$s_1, s_2$ of $\bH_\bC$, and a $C^\infty$ vector field $\xi$ on $\cM_B$, we have
$$
L_\xi \left( Q(s_1,s_2) \right) = Q(\nabla_\xi  s_1, s_2) + Q(s_1, \nabla_\xi s_2),
$$
where 
$$
L_\xi : C^\infty(\cM_B;\bC)\to C^\infty(\cM_B;\bC) 
$$
is the Lie derivative on $C^\infty$ functions, and where 
$$
\nabla_\xi: C^\infty(\cM_B,\bH_\bC)\to C^\infty(\cM_B, \bH_\bC)
$$
is the covariant derivative defined by the Gauss-Manin connection.

\subsection{Special homogeneous coordinates and the period map}
Let $s:\tcM_B\to p^*\cL^\vee$ be the tautological section. Write
$$
s(\xi) = \sum_{i=0}^k (x^i(\xi) \alpha_i + p_i(\xi) \beta^i).  
$$
Then $x^0(\xi),\ldots, x^\kappa(\xi)$ are the local holomorphic coordinates on the extended complex moduli $\tcM_B$
and the local homogeneous coordinates on the complex moduli $\cM_B$; they are called ``special homogeneous coordinates'' in \cite{S}.  

Let $V = \left(H^{\oplus (\kappa+1)}\right)\otimes_\bZ \bC$, which is a complex symplectic vector space of dimension $2(\kappa+1)$. 
The isometry $\gamma: H^{\oplus(\kappa+1)} \to H^3(X_q;\bZ)/\mathrm{Tor}$ extends to an isomorphism $\gamma: V\to H^3(X_q;\bC)$ of complex symplectic vector spaces.
There is a period map
$$
P_\cT: \cT \to \bP(V)\cong \bP^{2\kappa+1}, \quad (X_q,\gamma) \mapsto  \gamma^{-1}(H^{3,0}(X_q)).
$$
More explicitly, 
$$
P_\cT(q)= [x^0(\xi), x^1(\xi),\ldots, x^\kappa(\xi), p_0(\xi), p_1(\xi),\ldots, p_{\kappa}(\xi)],
$$
where $\xi \in \tcT$ is any point in the fiber of $\tcT\to \cT$ over $q\in \cT$.

Let $\cO_{\bP(V)}(-1)$ be the tautological line bundle over $\bP(V)$. Then 
$$
P_\cT^* \cO_{\bP(V)}(-1) = \pi^*\cL^\vee.
$$

Recall the Euler sequence:
$$
0\to \cO_{\bP(V)} \to  V\otimes \cO_{\bP(V)}(1) \to T_{\bP(V)} \to 0.
$$
Pulling back the above sequence under $P_\cT$, we obtain
$$
0\to \cO_\cT \to V \otimes \pi^*\cL \to  P_\cT^*T_{\bP(V)} \to 0.
$$
Here, $\pi: \cT \to \cM_B$ is the projection from the Torelli space to the 
complex moduli as before, and 
$$
V\otimes \pi^*\cL = \pi^*(\bH_\bC\otimes \cL). 
$$

\subsection{The Hodge metric}
The symplectic form $Q$ on $H^3(Y;\bR)$ extends to $V= H^3(Y;\bC)$. For $\alpha,\beta\in H^3(Y,\bC)$, define
$$
(\alpha,\beta):= \sqrt{-1}Q(\alpha,\beta).
$$

By the Hodge-Riemann bilinear relation,
\begin{itemize}
\item $(\alpha,\alpha) =0$ if $\alpha\in H^{3,0}(Y)\oplus H^{2,1}(Y)$;
\item $(\alpha,\bar{\alpha})>0$  if $\alpha\in H^{3,0}(Y)$ and $\alpha\neq 0$;
\item $(\alpha,\bar{\alpha})<0$  if $\alpha\in H^{2,1}(Y)$ and $\alpha\neq 0$.
\end{itemize}
Define a Hermitian metric on the holomorphic line bundle $\cL^\vee=\bF^3$ by 
$$
\parallel \Omega \parallel^2 = (\Omega,\bar{\Omega}).
$$
This is known as the Hodge metric on $\bF^3$. 

Let $\bH^{2,1}$ be the complex vector bundle over $\cM_B$ whose fiber over $q\in \cM_B$ is $H^{2,1}(Y_q)$. Then 
$\bH^{2,1}$ is a $C^\infty$ complex subbundle of $\bH_\bC$ but not a holomorphic subbundle of $\bH_\bC$.
Define the Hodge metric $h$ on $\bH^{2,1}$ by 
$$
h(s_1,s_2) = -(s_1,\bar{s}_2).
$$
Then $(\bH^{2,1}, h)$ is a Hermitian vector bundle of rank $\kappa$ over $\cM_B$.

\subsection{The Chern connection on $\cL^\vee = \bF^3$}\label{sec:chern}
Let 
$$
D:\Omega^0(\cM_B, \cL^\vee) \to \Omega^1(\cM_B, \cL^\vee) = \Omega^{1,0}(\cM_B,\cL^\vee) \oplus \Omega^{0,1}(\cM_B,\cL^\vee)
$$
be the Chern connection determined by the holomorphic structure and the Hodge metric on $\cL^\vee=\bF^3$, and let 
$D'$ and $D''$ be the $(1,0)$ and $(0,1)$ parts of $D$, so that
$$
D = D' + D'',
$$
where $D''$ depends on the holomorphic structure. Any $C^\infty$ section  $s$ of $\cL^\vee$ is also a $C^\infty$ section of $\bH_\bC$, and we have
$$
D''s = \nabla''s
$$
since $\cL^\vee$ is a holomorphic subbundle of $\bH_\bC$.  

If $(Y_q,\Omega_q)$ is a local holomorphic frame of $\cL^\vee$  over an open neighborhood $U$ then $D''\Omega =0$,  and for any tangent vector
$v\in T_q \cM_B$, where $q\in U$, we have 
\begin{equation} \label{eqn:Dv}
D'_v \Omega = D_v \Omega =  \frac{(\nabla_v \Omega, \bar{\Omega})}{(\Omega, \bar{\Omega})} \Omega.
\end{equation}
The connection 1-form is
$$
A= \partial \log(\Omega,\bar{\Omega}) \in \Omega^{1,0}(U).
$$
The curvature form 2-form is
$$
F = d A  = -\partial \bar{\partial} \log \left( \parallel \Omega \parallel^2\right) \in \Omega^{1,1}(U). 
$$
The right-hand side is independent of the choice of the local holomorphic frame, so $F$ is a global $(1,1)$ form on $\cM_B$.
More explicitly, we define
$$
\bD: \Omega^0(\cM_B,\cL^\vee)\to \Omega^{1,0}(\cM_B, \bH^{2,1})
$$
by 
$$
\bD s = \nabla' s - D's = \nabla s - D s.
$$
Write
$$
F = F_{i\bar{j}} dz^i \wedge d\bar{z}^j,
$$
where $(z^1,\ldots, z^\kappa)$ are the local holomorphic coordinates on $U\subset \cM_B$. Define
$$
\nabla_i:= \nabla_{\frac{\partial}{\partial z^i}}, \quad D_i:= D_{\frac{\partial}{\partial z^i}}, \quad 
\bD_i =  \nabla_i - D_i.
$$
Then 
\begin{eqnarray*}
F_{i\bar j} &=& -\frac{\partial^2}{\partial z^i \partial \bar{z}^j}  \log(\Omega,\bar{\Omega})\\
&=& -\frac{\partial}{\partial z^i} \frac{(\Omega, \overline{\nabla_j\Omega})}{(\Omega,\bar{\Omega})} \\
&=& -\frac{(\nabla_i \Omega, \overline{\nabla_j\Omega})}{(\Omega,\bar{\Omega})} +
\frac{ (\nabla_i\Omega, \bar{\Omega}) (\Omega, \overline{\nabla_j\Omega}) }{(\Omega,\bar{\Omega})^2} \\
&=&  \frac{-(\nabla_i\Omega,\overline{\nabla_j\Omega} ) + (D_i\Omega, \overline{D_j\Omega} )}{(\Omega,\bar{\Omega})} \\
&=& \frac{ -(\bD_i \Omega,\overline{\bD_j\Omega})  }{ (\Omega,\bar{\Omega}) }
\end{eqnarray*}
where the fourth equality follows from Equation \eqref{eqn:Dv}, and the fifth (and last) equality follows from 
the identity $\nabla_k \Omega = D_k \Omega + \bD_k\Omega$ where $D_k\Omega$ is a (3,0)-form and $\bD_k \Omega$ is a (2,1)-form.

The above computation shows that the first Chern form 
$$
c_1(\cL^\vee, D) = \frac{\sqrt{-1}}{2\pi} F  
$$
is a positive $(1,1)$-form.  

\subsection{The Weil-Petersson metric}
The Weil-Petersson metric on $\cM_B$ is defined by the  K\"{a}hler form  
$$
\omega_{\mathrm{WP}} := 2\pi c_1(\cL^\vee,D) = \sqrt{-1} F.
$$
In local holomorphic coordinates, the  Weil-Petersson metric is given by 
$$
G_{i\bar{j}} := \langle \frac{\partial}{\partial z_i}, \frac{\partial}{\partial z_j} \rangle
=  \frac{-(\bD_i\Omega,\overline{\bD_j\Omega})}{(\Omega,\bar{\Omega})} = \frac{g_{i\bar{j}}}{g_{0\bar{0}}},
$$
where $g_{i\bar{j}}$ is the Hodge metric on $\bH^{2,1}$ and $g_{0\bar{0}}$ is the Hodge metric on $\cL^\vee=\bF^3$.

We have an isomorphism of Hermitian vector bundles:
\begin{equation} \label{eqn:tangent-hodge}
T\cM_B \otimes \cL^\vee \cong  \bH^{2,1}, 
\end{equation}
where the tangent bundle $T\cM_B$ is equipped with the Weil-Petersson metric $G_{i\bar{j}}$, while
$\bH^{2,1}$ is equipped with the Hodge metric $g_{i\bar{j}}$, and 
$\cL^\vee$ is equipped with the Hodge metric $g_{0\bar{0}}$. 
The restriction of \eqref{eqn:tangent-hodge} to a point $q\in \cM_B$ can be identified with
$$
H^1(Y_q,T_Y)\otimes H^0(Y_q,\Omega^3_{Y_q}) \cong  H^1(Y_q,\Omega^2_{Y_q}). 
$$

\section{B-model on compact Calabi-Yau threefold}

\subsection{Genus zero free energy }
$$
x^0, x^1,\ldots, x^\kappa, p_0, p_1,\ldots, p_\kappa \in \Gamma(\cT, p^*\cL)
$$
are holomorphic sections of the line bundle 
$$
p^*\cL = \cP_{\cT}^*\cO_{\bP(V)}(1)
$$
over the Torelli space $\cT$. Note that
$$
\cF_0 := \frac{1}{2} \sum_{i=0}^\kappa x^i p_i
$$
is a holomorphic section on $p^*\cL^2 \to \cT$ and is a multi-valued holomorphic section of $\cL^2\to \cM_B$. 
Define the local holomorphic functions
\begin{equation}\label{eqn:Fi}
z^i:= \frac{x^i}{x^0},\quad F_i:= \frac{p_i}{x^0},\quad  g := \frac{p_0}{x^0},\quad \check{F}_0 = (x^0)^{-2} \cF_0,
\end{equation}
where $i=1,\ldots,\kappa$. In particular,
$$
g = 2\check{F}_0 -\sum_{i=1}^\kappa z^i F_i.
$$
The functions $(z^1,\ldots, z^\kappa)$ are known as special coordinates (or the B-model flat coordinates), 
which are  local  holomorphic coordinates on $\cT$, defined in an open
neighborhood of large complex structure with $A^0$ a vanishing cycle. The function $\check{F}_0$ is the B-model genus zero free energy. 

We may write $\Omega = x^0  \Omega_0$, where
$$
\Omega_0 =  \alpha_0 + \sum_{j=1}^\kappa z^j \alpha_j +  \sum_{j=1}^\kappa F_j \beta^j 
+ (2\check{F}_0 - \sum_{j=1}^\kappa z^j F_j)\beta^0. 
$$
For $i=1,\ldots,\kappa$, 
$$
\nabla_i \Omega_0 = \alpha_i + \sum_{j=1}^\kappa \frac{\partial F_j}{\partial z^i}\beta^j
+ \left(2\frac{\partial \check{F}_0}{\partial z^i} - F_i - \sum z^j\frac{\partial F_j}{\partial z^i}  \right)\beta^0.
$$
Here, $\Omega_0$ is a local holomorphic section of $\bF^3$, and $\nabla_i \Omega_0$ is a local holomorphic 
section of $\bF^2$ by Griffiths transversality, so Hodge-Riemann bilinear relations imply  
$$
0 = Q(\Omega_0, \nabla_i \Omega_0) = 2\frac{\partial \check{F}_0}{\partial z^i} - 2 F_i. 
$$

To summarize: 
\begin{itemize}
\item The B-model flat coordinates are given by 
$$
z^i = \frac{x^i}{x^0} =\frac{\int_{A^i}\Omega}{\int_{A^0}\Omega},\quad  i=1,\ldots, \kappa.
$$
\item The B-model genus zero free energy $\check{F}_0$ is defined by
$$
\check{F}_0 =\frac{1}{2} (x^0)^{-2}  \cdot  \sum_{i=0}^{\kappa} x^ip_i, 
$$
which satisfies 
$$
\frac{\partial \check{F}_0}{\partial z^i} = \frac{p_i}{x^0}=  
\frac{\int_{B_i}\Omega}{\int_{A^0} \Omega},\quad i=1,\ldots, \kappa.
$$
\end{itemize} 

\subsection{Yukawa coupling}
$$
\nabla_i \Omega_0 = \alpha_i + \sum_{j=1}^\kappa \frac{\partial^2 \check{F}_0}{\partial z^i\partial z^j} \beta^j \mod \beta^0. 
$$
$$
\nabla_j \nabla_k \Omega_0 = \sum_{j,k,\ell=1}^\kappa  
\frac{\partial^3 \check{F}_0}{\partial z^j\partial z^k \partial z^\ell} \beta^\ell
\mod \beta^0.
$$
$$
Q(\nabla_i \Omega_0, \nabla_j \nabla_k \Omega_0) = 
\frac{\partial^3 \check{F}_0}{\partial z^i \partial z^j\partial z^k}. 
$$
We also have
$$
Q(\nabla_i \Omega_0, \nabla_j \nabla_k \Omega_0) = \frac{\partial}{\partial z^i} Q(\Omega_0, \nabla_j\nabla_k \Omega_0)
-Q(\Omega_0, \nabla_i \nabla_j \nabla_k \Omega_0) = -Q(\Omega_0, \nabla_i \nabla_j \nabla_k \Omega_0).
$$
So
$$
\frac{\partial^3 \check{F}_0}{\partial z^i \partial z^j\partial z^k} =  -Q(\Omega_0, \nabla_i \nabla_j \nabla_k \Omega_0)
=-\int_Y \Omega_0 \wedge \nabla_i \nabla_j \nabla_k \Omega_0. 
$$
Define $C\in \Gamma((T^*M)^3\otimes \cL^2 )$ by 
\begin{eqnarray*}
C_{ijk} = C(\frac{\partial}{\partial z^i}, \frac{\partial}{\partial z^j}, 
\frac{\partial}{\partial z^k}) &:=& -\int_Y \Omega\wedge \nabla_i \nabla_j \nabla_k \Omega\\
&=& -(x^0)^2 \int_Y \Omega_0 \wedge \nabla_i \nabla_j \nabla_k \Omega_0 
= (x^0)^2 \frac{\partial^3 \check{F}_0}{\partial z^i \partial z^j\partial z^k}.
\end{eqnarray*}
Then $C_{ijk}$ is symmetric in $i,j,k$, so 
$$
C \in \Gamma(\Sym^3(T^*M)\otimes \cL^2 ).
$$

Define
$$
C_{\bar{i}}^{jk} := \overline{C_{ilm}}g^{j\bar{l}} g^{k\bar{m}}. 
$$
Then
\begin{equation}\label{eqn:C}
\cC:= C_{\bar{i}}^{jk}d\bar{z}^i \otimes \frac{\partial}{\partial z_j} \otimes
\frac{\partial}{\partial z_k} \in \Omega^{0,1}(\cM_B, (T\cM_B)^{\otimes 2} \otimes \cL^{-2}). 
\end{equation}

\subsection{Genus one free energy}
The genus one free energy 
$\check{F}_1$ is a linear combination of Ray-Singer torsions. A mathematical definition of $\check{F}_1$ is given in 
\cite{FLY}.

\subsection{Genus $g\geq 2$ free energies and the Holomorphic Anomaly Equations}
The Weil-Petersson metric is K\"{a}hler, so the Chern connection on $T\cM_B$ defined by the holomorphic structure and the 
Weil-Petersson metric is also torsion free. We equip
$\cL^\vee=\bF^3$ with the connection $D$ in Section \ref{sec:chern}. These two connections induce connections on tensor bundles
$$
(T^*\cM_B)^{\otimes m} \otimes \cL^k
$$
for any integers $m,k$. Let $D$ be the covariant derivative defined by these connections.

The special homogeneous coordinates $x^0,x^1,\ldots, x^{\kappa}$ are local holomorphic sections of the vacuum line bundle $\cL$. 
For $g\geq 2$,
$$
\cF_g(z,\bar{z}) \in \Gamma(\cM_B, \cL^{2-2g}).
$$
The limit
$$ 
\check{F}_g(z)=\lim_{\bar{z}\to \sqrt{-1}\infty} (x^0)^{2g-2}\cF_g(z,\bar{z})
$$
is a holomorphic function on $U$. The non-holomorphic section $\cF_g(z, \bar{z})$ satisfies the following Holomorphic Anomaly Equation (BCOV \cite{BCOV}):
\begin{equation}\label{eqn:BCOV}
\dbar_i \cF_g =  \frac{1}{2} C_{\bar{i}}^{jk}  \Big(D_j D_k \cF_{g-1} + \sum_{\substack{g_1,g_2>0\\ g_1+g_2=g} } D_j \cF_{g_1}D_k \cF_{g_2} \Big).
\end{equation}

More precisely, we have
$$
DD\cF_{g-1} + \sum_{\substack{g_1,g_2>0\\ g_1+g_2=g} } D\cF_{g_1} \otimes D\cF_{g_2}  \in  \Gamma( \cM_B, (T^*\cM_B)^{\otimes 2}\otimes \cL^{4-2g}),
$$
$$
 \cC  \in  \Omega^{0,1}(\cM_B, (T\cM_B)^{\otimes 2} \otimes \cL^{-2}),
$$
where $\cC$ is defined by Equation \eqref{eqn:C}. Using the natural pairing between $T\cM_B$ and $T^*\cM_B$, we obtain 
$$
\cC \diamond \Big( DD\cF_{g-1} + \sum_{\substack{g_1,g_2>0\\ g_1+g_2=g} } D\cF_{g_1} \otimes D\cF_{g_2}\Big) \in \Omega^{0,1}(\cM_B, \cL^{2-2g}).
$$
With the above notation, the Holomorphic Anomaly Equation \eqref{eqn:BCOV} can be rewritten in the following coordinate-free form:
\begin{equation}\label{eqn:BCOV-tensor}
D''\cF_g = \frac{1}{2} \cC \diamond \Big( DD\cF_{g-1} + \sum_{\substack{g_1,g_2>0\\ g_1+g_2=g} } D\cF_{g_1} \otimes D\cF_{g_2}\Big).
\end{equation}

\subsection{B-model topological open string and the Extended Holomorphic Anomaly Equation}
Let 
$$
\Delta_{ij} := D_jD_i \cF_{(0,1)} 
$$
be the second covariant derivatives of the disk potential, and define
$$
\Delta_{\bar{i}}^j := g^{j\bar{k}} \overline{\Delta_{ki}}. 
$$
Then 
$$
\Delta_{\bar{i}}^j d\bar{z}^i \otimes \frac{\partial}{\partial z_j} \in \Omega^{0,1}(\cM_B, T\cM_B \otimes \cL^{-1}). 
$$
If $h>0$ and $2g-2+h>0$, then
$$
\cF_{(g,h)}(z,\bar{z}) \in \Gamma(\cL^{2-2g-h}).
$$
The limit
$$
\check{F}_{(g,h)}(z)=\lim_{\bar{z}\to \infty} (x^0)^{2g-2+h}\cF_{(g,h)}(z,\bar{z})
$$
is a holomorphic function on $U$. The non-holomorphic section
$\cF_{(g,h)}(z,\bar{z})$ satisfies the following Extended Holomorphic Anomaly Equation (Walcher \cite{W})
\begin{equation}\label{eqn:Walcher}
\bar{\partial}_i \cF_{(g,h)} =  \frac{1}{2} C_{\bar{i}}^{jk}  \Big(D_j D_k \cF_{g-1,h} +
 \sum'_{\substack{g_1+ g_2 =g\\ h_1+h_2=h} } D_j \cF_{g_1,h_1}D_k \cF_{g_2,h_2} \Big) -\Delta_{\bar{i}}^j D_j F_{g,h-1},
\end{equation}
where the sum $\displaystyle{\sum'}$ excludes the unstable case $(g_i,h_i)=(0,0), (0,1)$, and the last term on the RHS corresponds to the cases 
$(g_1,h_1)=(0,1)$ or $(g_2,h_2)=(0,1)$.

\subsection{Mirror symmetry}
Under the mirror map
$$
(t^1,\ldots, t^\kappa) = (z^1(q),\ldots, z^\kappa(q))  
$$
we have the following mirror conjectures:
\begin{itemize}
\item For any $i,j,k\in \{1,\ldots, \kappa\}$, 
$$
\frac{\partial^3\check{F}_0}{\partial z^i \partial z^j \partial z^k}(z)
=\frac{\partial^3 F_0^X}{\partial t^i \partial t^j \partial t^k}(t).
$$
\item For $i=1,\ldots, \kappa$, 
$$
\frac{\partial \check{F}_1}{\partial z^i}(z) = \frac{\partial F_1^X}{\partial t^i}(t),
$$
or equivalently,
$$
d\check{F}_1  = d F_1^X.
$$

\item For $g\geq 2$,
$$
\check{F}_g(z) = F_g^X(t). 
$$

\item If $h>0$ and $2g-2+h>0$ then
$$
\check{F}_{(g,h)}(z) = F_{(g,h)}^{X,L}(t). 
$$ 
\end{itemize}

\subsection*{Acknowledgments} This note is based on the author's talks at
the 6th Workshop on Combinatorics of Moduli Spaces, Cluster Algebras, and Topological Recursion
in Moscow on June 4--9, 2018, and the conference on Crossing the Walls in Enumerative Geometry
in Snowbird, Utah on May 21--June 1, 2018. The author sincerely thanks the organizers of these events
for the invitation to participate as a speaker. The author also wishes to thank Bohan Fang, Sheldon Katz, Zhengyu Zong, and the anonymous referee for their helpful comments on an earlier version of this note. 

\end{document}